\documentclass[conference,compsoc]{IEEEtran}
\ifCLASSOPTIONcompsoc
  \usepackage[nocompress]{cite}
\else
  \usepackage{cite}
\fi
\ifCLASSINFOpdf
\else
\fi
\usepackage{amssymb}
\usepackage{graphicx}
\usepackage{enumitem}
\usepackage{pgfplots}
\usepackage{xurl}
\pgfplotsset{compat=1.17}
\usepackage{subcaption}
\usepackage{csvsimple}
\usepackage{siunitx}
\usepackage{tabularx}
\usepackage{colortbl}
\usepackage{xcolor}
\usepackage{float}
\usepackage{tikz}
\usepackage{hyperref}
\usepackage{tcolorbox}
\usetikzlibrary{shapes, positioning}
\usepackage{tikz}
\usetikzlibrary{positioning, shapes, arrows.meta}
\usepackage{todonotes}

\usepackage{pifont}

\begin{document}
%
\title{A Study of Malware Prevention in Linux Distributions}

\author{\IEEEauthorblockN{Duc-Ly Vu}
\IEEEauthorblockA{ 
Eastern International University\\
ly.vu@eiu.edu.vn}
\and
\IEEEauthorblockN{Trevor Dunlap}
\IEEEauthorblockA{Chainguard\\
trevor.dunlap@chainguard.dev}
\and
\IEEEauthorblockN{Karla Obermeier-Velazquez}
\IEEEauthorblockA{Purdue University \\
obermeier.velazquez@gmail.com}
\and
\IEEEauthorblockN{Thanh-Cong Nguyen}
\IEEEauthorblockA{University of Information Technology\\
thanh-congn@acm.org}
\and
\IEEEauthorblockN{Paul Gilbert}
\IEEEauthorblockA{Chainguard \\
paul.gibert@chainguard.dev}
\and
\IEEEauthorblockN{John Speed Meyers}
\IEEEauthorblockA{Chainguard \\
jsmeyers@chainguard.dev}
\and
\IEEEauthorblockN{Santiago Torres-Arias}
\IEEEauthorblockA{Purdue University \\
santiagotorres@purdue.edu}
}

\maketitle

\begin{abstract}
Malicious attacks on open-source software packages are a growing concern. 
The discovery of the \textit{XZ} Utils backdoor intensified these concerns because of the potential widespread impact.
This study, therefore, explores the challenges of preventing and detecting malware in Linux distribution package repositories. To do so, we ask two research questions: (1) What measures have Linux distributions implemented to counter malware, and how have maintainers experienced these efforts? (2) How effective are current malware detection tools in identifying malicious Linux packages? To answer these questions, we conduct interviews with maintainers at several major Linux distributions and introduce a Linux package malware benchmark dataset. Using this dataset, we evaluate the performance of six open-source malware detection scanners. Distribution maintainers, according to the interviews, have mostly focused on reproducible builds to date. Our interviews identified only a single Linux distribution, Wolfi OS, that performs active malware scanning. Using this new benchmark dataset, the evaluation found that the performance of existing open source malware scanners is underwhelming. Most studied tools excel at producing false positives but only infrequently detect true malware. Those that avoid high false positive rates often do so at the expense of a satisfactory true positive. Our findings provide insights into Linux distribution package repositories' current practices for malware detection and demonstrate the current inadequacy of open source tools designed to detect malicious Linux packages.
\end{abstract}


%
\IEEEpeerreviewmaketitle

\section{Introduction}

When GitHub user \textit{JiaT75} made their first commit to the open-source library \texttt{XZ} in 2021, it was uneventful~\cite{xz}.
It was just another day and another commit. When this user later became a maintainer of the XZ Utils library, a widely used open-source software library for data compression, it was still uneventful. It was another case of a strapped open-source project adding a new maintainer.
These seemingly routine events drew widespread attention in 2024 after a Microsoft engineer discovered that \textit{JiaT75} had then used their maintainer privileges to insert a backdoor into two released versions of \texttt{XZ} Utils~\cite{goodin2024}.
The revelation that an individual or group masqueraded as a typical open-source software maintainer of the popular \texttt{XZ} Utils library solely to implant malicious code in it has caused widespread concern among observers of software security and open-source software.

Malicious open-source software packages are not a new phenomenon. There have been thousands of malicious packages in popular open-source package registries~\cite{duan2021towards, ohm2020backstabber}.
These instances, however, often involve attack methods like typosquatting, an attacker tricking a victim into downloading the ``wrong" open-source software package, e.g., \texttt{reqeusts} rather than \texttt{requests}, not the insertion of backdoors in popular packages~\cite{vu2020typosquatting}.
There are, admittedly, known instances of open-source software maintainers who have turned malicious. 
In one widely reported incident, a maintainer abruptly removed a popular npm library, temporarily breaking countless web applications~\cite{its_2016}. Another maintainer of the popular \texttt{event-stream} library inserted code to harvest credentials from Bitcoin wallets~\cite{arvanitis_2022}. 
However, these maintainer-related instances typically involve either a compromise of availability, which is arguably less severe than a compromise of confidentiality or integrity, or they target a limited set of users. 
These malicious packages were all found in package registries where the barriers to publishing software are intentionally low, in registries, such as the Python Package Index (PyPI), npm, and RubyGems.

The \texttt{XZ} Utils compromise was different. If the malicious versions had gone undiscovered for longer and had been incorporated into the stable branches of major Linux distributions, the integrity and confidentiality of the systems of many users could have been affected. In one participant’s evocative phrase, this attack, if successful, would have endowed the attackers with a ``skeleton key" for the Internet~\cite{Nitansha_2024}. In other words, a malicious package in a major Linux distribution, in comparison to yet another typosquatted open-source software package, is a different class of problem. 

The \texttt{XZ} Utils compromise raises important questions that have not yet been thoroughly addressed.
At the broadest level, what should be done to prevent and detect malware from entering the package registry of Linux distributions? More practically, what is feasible and desirable in the short term? To help answer aspects of these broad questions, we set out to answer two specific research questions:
\begin{itemize}
    \item \textbf{RQ1}: Based on interviews with maintainers at several Linux distributions, what have Linux distributions done previously to counter malware? What has the experience of these maintainers been with these approaches to countering malware? 
    \item \textbf{RQ2}: Based on a novel Linux package malware benchmark dataset, what is the performance of existing source code and Linux package binary malware detection tools?
\end{itemize}

\textit{Our broader goal is to provide an analytical basis for maintainers at major Linux distributions to make an informed decision about the benefits and costs of different approaches to detecting Linux package malware, especially approaches that rely on proactive scanning.} To achieve the goal, we make the following contributions:
\begin{itemize}
    \item The interview portion of the study provides an in-depth understanding of the perceptions and experiences of maintainers at major Linux distributions trying to prevent and detect malware from being uploaded to their package repositories.
    \item A set of six new Linux package malware datasets that can be used to assess the performance of tools that aim to detect malware in Linux distributions
    \item An assessment of malware detection scanners and one capability analysis tool on the benchmark dataset.

\end{itemize}

To the best of our knowledge, this is the first human-aspects interview protocols with Linux Distribution contributors on this topic.
Similarly, while previous literature has explored the sensitivity and ROC-curve performance of off-the-shelf antivirus on OSS malware, the data collected complements the insights provided by our interview protocol.
In combination, these findings shed light on the shed light on a fundamental gap on these tools.
We believe that our dataset is a foundational step in providing a baseline for new tools to detect and prevent software supply chain attacks on the Linux Distribution ecosystem.

\section{Background}
\label{sec:background}
\subsection{A Distinct Software Supply Chain Security Sub-Field Has Emerged}
\noindent
This study fits within a broader stream of research on software supply chain security, an analytical agenda that focuses on the security of the code, tools, and processes that underpin the building and deployment of a piece of software. While this area of inquiry has existed for decades~\cite{thompson1984reflections, wheeler2005countering, lamb2021reproducible, torres2019toto, geer2020good}, recent high-level compromises and vulnerabilities have thrust this topic into the mainstream. Notably, during the SolarWinds hack, the Russian intelligence services broke into the build system of the network management software company SolarWinds and implanted malware into the software updates distributed to its customers~\cite{Office_2022}. Governments, companies, funding agencies, and researchers have all taken note~\cite{cisa, zahan2024s3c2}.

Consequently, there has been an explosion of activity related to software supply chain security. One prominent topic has been software bill of materials (SBOMs), the software equivalent of an ingredients list, which advocates believe can provide transparency into the components of a piece of software and ultimately be used to improve security for both producers and consumers of software~\cite{xia2023empirical, zahan2023sbom}. Software signing has also been a key topic. Systems like Sigstore offer the possibility of increasing the adoption of software signatures so that consumers of software can check the validity of these signatures, ensuring that any tampering is caught before the software is consumed~\cite{newman2022sigstore, merrill2023speranza}. Software provenance has also been featured prominently. For instance, Supply-Chain Levels for Software Artifacts (SLSA) has become a popular analytical framework used to assess provenance~\cite{slsa}.

\subsection{Proactive Scanning for Malicious Open Source Packages Has Become Popular}
\noindent
One area of particular relevance has been an increase in research on proactively detecting malicious open-source software packages. One pioneering project, for instance, found over three hundred malicious open-source software packages on PyPI, npm, and RubyGems~\cite{duan2021towards}. Malicious code can be injected in a compiler~\cite{thompson1984reflections}, the build, or in continuous integration systems~\cite{moriconi2023reflections}. Many commercial companies have claimed to offer tools capable of detecting malicious open-source software packages, often self-publishing articles that describe how their company’s tool detected malicious packages on popular registries~\cite{vu2023bad}.  There has also been academic research on the topic, often emphasizing machine learning approaches to malicious open source package detection~\cite{ohm2022feasibility, zahan2024shifting}. The study by Vu et al.~\cite{vu2023bad} examined available tools for detecting malicious Python packages and found that the tools had extremely high false positive rates, so high that repository administrators found the tools unsuitable for their own use.

These papers on proactive open-source package malware scanning have almost exclusively focused on programming language registries like PyPI, npm, RubyGems, and others. This focus has been reasonable. After all, the vast majority of malicious open-source software packages have been found in these registries~\cite{ohm2020backstabber}. But the \textit{XZ} Utils attack, which targeted major Linux distributions, suggests that open-source software package malware researchers should not exclusively focus on programming language registries.

\subsection{Linux Distributions Are Also Worthy of the Attention of Software Supply Chain Security Researchers}
\noindent
In the literature, some researchers have done research on Linux distributions and software supply chain security. Notably, the \textit{Reproducible Builds} project has focused on Debian~\cite{lamb2021reproducible}, but the majority of software supply chain security research related to open-source software has focused on programming language package registries~\cite{ladisa2023sok}. This omission is concerning and not only because of the \textit{XZ} Utils attack, but also because the software from Linux distributions is arguably just as widespread and heavily used as the software from major programming language registries. Moreover, the defenses of Linux distributions to malware are relatively under-examined, more often assumed than analyzed, let alone improved based on careful analysis. This is why this research set out to first understand the current state of approaches to countering malware in several major Linux distributions and then assess several tools that could potentially be used to detect malware before it enters a Linux distribution package repository.
\section{Methodology}
\label{sec:methods}
The methodologies of this research can be divided into two parts. To answer the first research question (\textbf{RQ1}), which examines how maintainers at major Linux distributions counter malware, this paper employed an interview methodology, selecting and interviewing key maintainers from a range of major Linux distributions. To answer the second research question (\textbf{RQ2}), we used a two-step method: (1) the research team first built six distinct datasets appropriate for benchmarking any tool that purports to detect Linux package malware and (2) the team subsequently used these benchmark datasets to evaluate the performance of various open-source tools. 


\subsection{Interviewing Maintainers at Major Linux Distributions about Countering Malware}
\noindent
This study used a qualitative, interview-based user research approach to understand the approaches maintainers at major Linux distributions use to counter malware and their perspectives on approaches they have employed or considered employing. This approach was chosen because knowledge about Linux distributions and malware is largely unavailable through traditional academic outlets, with the exception of a strain of research on Reproducible Builds \cite{lamb2021reproducible}. There is, to our knowledge, no prior work on proactive malware scanning and the perspective of Linux maintainers on this topic. Furthermore, what limited written knowledge that does exist is instead mostly scattered throughout email lists or IRC chats. Because key maintainers often possess insights and experiences that are otherwise inaccessible to outsiders, the study methods focused on interviews with these maintainers.

\begin{table}[h]
\centering
\caption{Interviewees}
\label{table:interviewees}
\setlength{\tabcolsep}{22pt}
\begin{tabular}{ccc}
\hline
\# & Distro & Years of Experience\\ \hline
\rowcolor[HTML]{EEEEEE} 1  &  Alpine & 20   \\
2  &  Arch & 8   \\ 
\rowcolor[HTML]{EEEEEE} 3  &  Arch & 20   \\ 
4  &  Debian & 20   \\ 
\rowcolor[HTML]{EEEEEE} 5  &  Debian & 25   \\ 
6  &  Ubuntu & 2   \\ 
\rowcolor[HTML]{EEEEEE} 7  &  Wolfi & 2   \\ 
\end{tabular}
\end{table}

The interviewees and their email addresses were identified through online searches for maintainers of these distributions and through introductions via existing personal contacts.
Table~\ref{table:interviewees} presents the key demographics of the interviewees in this study. All interviewees possessed extensive personal experience with the Linux distribution they represented in the interviews. Several interviewees had served as leaders or founders of their respective projects.

The seven interviewees were drawn from five different Linux distributions: \textit{Alpine}, \textit{Arch}, \textit{Debian}, \textit{Ubuntu}, and \textit{Wolfi}. There were two interviewees associated with Debian and two with Arch. Four of these five distributions were selected because they are long-running and widely used.  In this sense, they are meant to represent, from a methodological standpoint, a typical popular Linux distribution. The selection of \textit{Wolfi}, a relatively recent distribution, is the sole exception. This decision can be attributed to the research team’s affiliation (three members are affiliated with the company that maintains this Linux distribution) and the distribution’s proactive use of malware scanning, which makes it an outlier and a novelty. It is important to note that other major Linux distributions exist, and future researchers should consider extending this research to include them. 

All participants agreed to participate in the interviews. Most of the participants consented to being recorded to assist the research team. For the remaining interviews, the research team relied on detailed note-taking. The interview questions were designed to be straightforward. The interviews lasted between thirty minutes and one hour and were conducted via video-conference calls. We conducted the interviews following the ``semi-structured" format~\cite{yin2015qualitative}: The interviewer referred to the list of prepared questions but allowed the conversation to naturally explore unexpected topics as appropriate. The interview questionnaire included the following questions:
\begin{enumerate}[leftmargin=0.5cm]
    \item  Please tell us about yourself.
    \item How long have you had a relationship with Linux distro X?
    \item What is your relationship, past and present, with Linux distro X?
    \item Is ``malware” in distro X a problem you or other maintainers have been concerned about? If so, in what ways?
    \item What anti-malware techniques were either employed or considered to be employed at Linux distro X? 
    \item For those that were considered but not deployed, why?
    \item For those that were deployed, why were they chosen and what were the results?
    \item If you had advice for other distros on the topic of preventing malware from entering the distro, what would it be and why?
\end{enumerate}

\subsection{Building Benchmark Datasets}
\begin{figure}[t]
    \centering
    \begin{tikzpicture}[
        font=\scriptsize,  
        node distance=1.0cm and 1.6cm, 
        box/.style={rectangle, rounded corners, draw, align=center, minimum width=1.4cm, minimum height=0.6cm},
        dataset/.style={box, fill=white},
        source/.style={box, fill=white},
        scanner/.style={box, fill=white},
        arrow/.style={-{Stealth}, thick}
    ]

    \node[source] (backstabber) {Backstabber's\\Knife Collection};
    \node[dataset, right=of backstabber, yshift=1cm] (dataset1) {Dataset 1};

    \node[source, below=of backstabber, yshift=-0.3cm] (linux_source) {Linux source\\on The Wild};
    \node[dataset, right=of backstabber, yshift=-1cm] (dataset2) {Dataset 2};

    \node[source, below=of linux_source] (wolfi_apks) {Wolfi APKs};
    \node[source, right=of wolfi_apks] (upstream_repos) {Upstream repos};
    \node[dataset, below=of upstream_repos] (dataset4) {Dataset 3};
    \node[dataset, right=of linux_source] (dataset5) {Dataset 5};
    \node[dataset, right=of dataset4] (dataset3) {Dataset 4};

    \node[scanner, right=of dataset2] (apk_scanners) {APK\\scanners};
    \node[scanner, right=of dataset1] (source_code_scanners) {Source code\\scanners};

    \draw[arrow] (backstabber) -- node[midway, above left, align=center] {\scriptsize Select npm,\\PyPI, RubyGems} (dataset1);
    \draw[arrow] (dataset1) -- node[midway, left] {\scriptsize Create APKs} (dataset2);
    \draw[arrow] (dataset1) -- (source_code_scanners);
    \draw[arrow, dashed] (dataset2) -- (apk_scanners);
    \draw[arrow] (linux_source) -- node[midway,  align=center] {\scriptsize Create \\APKs} (dataset5);
    \draw[arrow] (wolfi_apks) -- node[midway,  align=center] {\scriptsize Extract\\ upstream} (upstream_repos);
    \draw[arrow, dashed] (wolfi_apks) to[out=45,in=240] (apk_scanners);
    \draw[arrow] (upstream_repos) -- node[midway, left, align=center] {\scriptsize Select repos,\\inject code} (dataset4);
    \draw[arrow] (upstream_repos) to[out=45,in=180] (source_code_scanners);
    \draw[arrow] (dataset4) -- node[midway,  align=center] {\scriptsize Create \\APKs} (dataset3);
    \draw[arrow] (dataset4) to[out=25,in=190] (source_code_scanners);
    \draw[arrow] (dataset3) -- (apk_scanners);
    \draw[arrow, dashed] (dataset5) -- (apk_scanners);

    \end{tikzpicture}
    \caption{Our Benchmarking of Malware Detection Tools}
    \label{fig:benchmarking-method}
\end{figure}
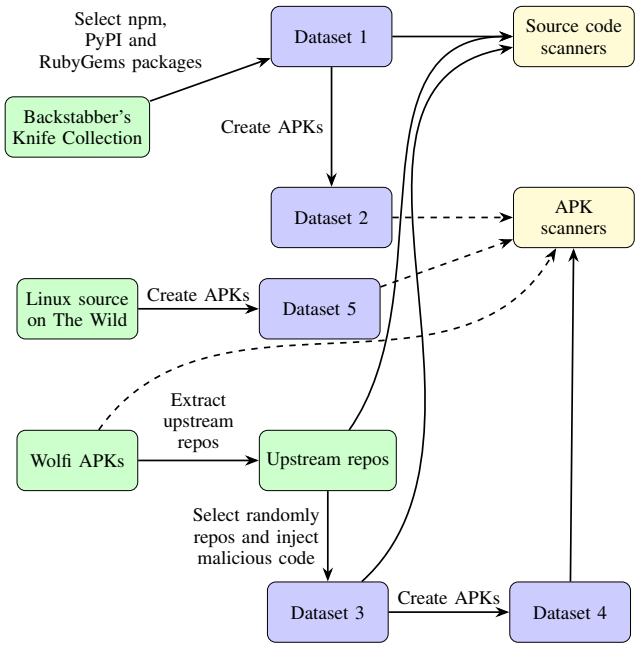

\noindent
The second methodological stage shown in Figure~\ref{fig:benchmarking-method} entailed constructing six distinct open-source Linux package malware datasets. These datasets, described in detail below, enabled the research team to precisely measure the performance of various tools. 

Before describing the datasets in detail, it is important to outline the methodological philosophy underpinning their design. First, as any individual dataset for benchmarking inherently has limitations, we sought to build a range of datasets that compensate for each other’s weaknesses. Second, each of the six malicious datasets has a corresponding benign dataset to enable benchmarking to measure both false positive and false negative rates. Third, some datasets focus on package binaries, specifically the APK format used by the Alpine and Wolfi Linux distributions, while others focus on source code. This decision reflects the project team’s uncertainty regarding the relative effectiveness of binary-based and source code-based approaches to detecting malicious Linux packages. Fourth, some types of open-source Linux package malware are arguably easier to detect (such as known examples of malware binaries repackaged into a Linux package), while others are harder to detect (such as novel malware designed for new attack vectors). To address this variability, we constructed multiple datasets along this spectrum of difficulty. Fifth, we included at least one dataset containing sequential package versions for the same package, with one version being malicious. This type of ``over-time" dataset enables this study and future research to evaluate tools that analyze cross-version differences, an approach employed by some open-source package malware detection tools.

\subsection{Benchmark Datasets}\label{subsec:datasets}

\subsubsection{Benign Datasets}
To construct a comprehensive dataset, we initially curated a set of benign examples from the Wolfi OS ecosystem~\cite{wolfi}. Each package in Wolfi is assumed to be benign. A total of \num{1866} Wolfi projects were selected, spanning the programming languages Python, JavaScript, Ruby, and C. Parsing the build files within Wolfi enabled the identification of upstream repositories for use with source code scanners. Each package includes an associated APK, which can be downloaded from the Wolfi APK index. The GitHub API~\cite{GitHubAPI} was used to identify the programming language of each upstream source code repository; only repositories in Python, JavaScript, Ruby, or C were selected for analysis. Additionally, the latest version available for each Wolfi package was utilized. Similarly, the latest commit of each upstream repository was checked out for use in source code analysis.

\begin{table*}[t]
    \centering
    \caption{ Datasets statistics}
    \begin{tabular}{llp{8cm}l}
        \hline
        Dataset Name & Type & Description & \# Samples \\
        \hline
        \rowcolor[HTML]{EEEEEE} Dataset \#1 & Source tarballs & Historical Samples of Open Source Source Code Malware & 30 \\
        Dataset \#2 & Wolfi APK & Historical Examples of Malicious Linux Binaries & 30 \\
        \rowcolor[HTML]{EEEEEE} Dataset \#3 &  Source tarballs & Synthetic Examples of Open Source Source Code Malware & 30 \\
        Dataset \#4 & Wolfi APK  & Synthetic Examples of Open Source Linux Binaries & 30 \\
        \rowcolor[HTML]{EEEEEE} Dataset \#5 & Wolfi APK & Synthetic Example of Linux Malicious Source Code turned into APKs & 10 \\
        Dataset \#6 & Wolfi APK & Synthetic Examples Over Time Golang Malware  & 10 \\
        \rowcolor[HTML]{EEEEEE} Wolfi Upstream & Source tarballs & Upstream repositories of Wolfi APKs & 1866 \\
        Wolfi APKs & Wolfi APK & Wolfi APKs & 1652 \\
        \hline
    \end{tabular}
    
    \label{tab:datasets}
\end{table*}

\subsubsection{Malicious Datasets}
In total, six distinct malware datasets were created, as summarized in Table~\ref{tab:datasets}. The methodological steps for creating each data set are shown in Figure~\ref{fig:benchmarking-method} and described in detail below:


\paragraph{\textit{Dataset \#1: Historical Malicious Examples of Open-Source Source Code}} We randomly selected a subset of malicious code samples from the Backstabber’s Knife Collection dataset, including ten Python projects, ten npm projects, and ten RubyGems projects~\cite{ohm2020backstabber}. 
In detail, a random sample was initially selected and compared with other samples in the dataset to determine whether they originated from the same attacks or shared the same behaviors. For example, Ohm et al.~\cite{ohm2020backstabber} shows that most packages in the dataset aim at data exfiltration. The study by Wang et al.~\cite{wang2025malpacdetector} shows that the Backstabber’s Knife Collection dataset has a significant percentage of duplications in malicious packages (\qty{77.3}{\percent}).
Our selection resulted in \num{30} packages, which form a dataset of ``historical" source code malware samples. To ensure that the sample included a range of attack strategies, we reviewed the malware techniques employed in each language ecosystem and ultimately made \num{15} replacements. 

\paragraph{\textit{Dataset \#2 - Malicious Linux binaries generated from known examples of malicious open source code}} The thirty samples selected in \textit{Dataset \#1} were then compiled into Wolfi APKs using \textit{melange}, Wolfi’s standard package build tool~\cite{melange}. Given a Wolfi APK, we follow this process: 
\begin{enumerate} 
    \item Extract its upstream source code repository 
    \item Identify the language used in the upstream source code based on the GitHub API. 
    \item Identify the injection locations. These locations had to be reachable during code execution.  
    \item Inject malicious code at the given locations. 
\end{enumerate} 
The injected malicious code was sourced from known examples. The injected code could reside in a main function that serves as the program's entry point or in an init function called during package initialization. Specifically, malicious code snippets were extracted from the datasets used in Vu et al.~\cite{vu2023bad}. We used malicious behaviors similar to those from go-malware, as seen in the malicious packages studied in~\cite{vu2023bad}. Examples of simple malicious behaviors included connecting to a remote server or collecting user information.

\paragraph{\textit{Dataset \#3 - Synthetic Source Code Malware}}
There are four steps to create this dataset:
\begin{enumerate} 
\item Take the upstream source code URL of a given Wolfi package
\item Determine the language used in the upstream source code
\item Identify injection locations. The locations must be reached during code execution, same as in Dataset \#2 
\item Inject malicious code at the injection location. 
\end{enumerate}

The injected malicious code is taken from known malicious examples present in the datasets used by Vu et al.~\cite{vu2023bad}. 
In particular, we injected simple malicious behaviors (e.g., those from go-malware), such as the malicious packages studied in~\cite{vu2023bad}. Examples of simple malicious behaviors include sending users’ information to a remote server.

\paragraph{\textit{Dataset \#4 - Synthetic examples of malicious Wolfi binaries}} The source code projects in \textit{Dataset \#3} were then packaged into APKs to enable analysis of binary malware detection tools. Building upon the process described in \textit{Dataset \#3}, we add two additional steps: 
\begin{enumerate} 
    \item Update the \textit{expected-commit} field in the \texttt{melange} YAML file to ensure the version of the source code with injected malware was built. 
    \item Build the APK from the updated upstream source code using \textit{melange}.
\end{enumerate}

\paragraph{\textit{Dataset \#5 - Synthetic example of Linux malware source code turned into binaries}} This dataset contained “full-fledged” malicious APKs. Malicious Linux source code projects targeting Linux distributions were selected from different sources, such as those cited in \cite{linux-malware}. Only source code targeting the x64 architecture was chosen. Ten malicious Wolfi APKs for this dataset were generated through the following steps: 
\begin{enumerate} 
    \item Select the source code of the malware. 
    \item Extract the build or compilation information from the source code. 
    \item Write and use the \textit{melange} YAML file for generating Wolfi APKs based on the extracted build information. 
\end{enumerate}

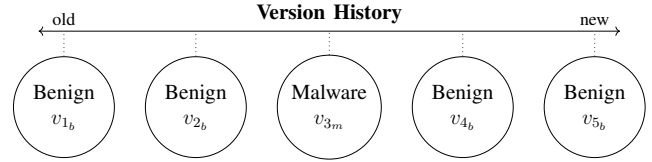
\begin{figure}[t]
    \centering
    \resizebox{\columnwidth}{!}{
        \begin{tikzpicture}
            \tikzstyle{version} = [circle, draw, minimum size=2cm, align=center, font=\large]
            
            \node[version] (v1) at (0, 0) {Benign\\$v_{1_b}$};
            \node[version, right=0.6cm of v1] (v2) {Benign\\$v_{2_b}$};
            \node[version, right=0.6cm of v2] (v3) {Malware\\$v_{3_m}$};
            \node[version, right=0.6cm of v3] (v4) {Benign\\$v_{4_b}$};
            \node[version, right=0.6cm of v4] (v5) {Benign\\$v_{5_b}$};

            \draw[dotted] (v1.north) -- ++(0,0.5) coordinate (v1top);
            \draw[dotted] (v2.north) -- ++(0,0.5);
            \draw[dotted] (v3.north) -- ++(0,0.5) coordinate (v3top);
            \draw[dotted] (v4.north) -- ++(0,0.5);
            \draw[dotted] (v5.north) -- ++(0,0.5) coordinate (v5top);

            \draw (v1top) -- (v5top);
            \draw[<-] (v1top) ++(-0.5,0) -- (v1top);
            \draw[->] (v5top) -- ++(0.5, 0);
            
            \node[anchor=south, font=\large] at (v3top) {\textbf{Version History}};
            \node[anchor=south] at (v1top) {\text{old}};
            \node[anchor=south] at (v5top) {\text{new}};
        \end{tikzpicture}
    }
    \caption{Construction of the ``over time'' dataset for a single project}
    \label{fig:dataset6-construction}
\end{figure}

\paragraph{\textit{Dataset \#6 - Synthetic example of over time Go malicious source code and APKs}} To identify malware, a common technique is to analyze changes in alerts generated by malware detection tools “over time” across different software versions. To simulate this process, we created a synthetic dataset of Go-based projects and injected malware from the Coldfire Golang malware development library~\cite{coldfire} into historical versions of these projects. We selected ten projects, each with a unique malware example. For example, one project was modified to execute shellcode directly in memory, while another was engineered to launch fork bombs, resulting in denial-of-service attacks. Each injected malware sample was validated to ensure it was reachable within the project's codebase and could be triggered to execute malicious behavior. To construct this dataset, we perform the following steps:
\begin{enumerate}
    \item Select the five latest versions ($V$) of an upstream Go Wolfi project. Go-based projects can be identified based on the build pipeline within the melange yaml file.
    \item Identify the five latest versions of the project, defined as $V=\{v_{1_b}, v_{2_b}, v_{3_b}, v_{4_b}, v_{5_b}\}$. As previously mentioned, we assumed all Wolfi projects and their versions to be benign.
    \item Inject malware into the middle version, $v_{3_b}$. Specifically, the set of over time dataset for a single project will be as follows: $V=\{v_{1_b}, v_{2_b}, v_{3_m}, v_{4_b}, v_{5_b}\}$, where $v_{3_m}$  is the modified malicious version replacing the benign version $v_{3_b}$. Figure~\ref{fig:dataset6-construction} shows the final construction of an over time dataset for a single project.
    \item Write and use the \textit{melange} YAML file for generating Wolfi APKs.
\end{enumerate}

\begin{table*}[!htb]
    \centering
    \caption{Selection criteria for Malware Source Code Detection Tools}
    \begin{tabular}{lccccc}
        \hline
        Malware Detection Tool & Open source & Malware detection & Multi-language support & Detection rules available & Capability Analysis\\
        \hline
        \rowcolor[HTML]{EEEEEE} Bandit4Mal & \checkmark & \checkmark &  & \checkmark & \\
        Capslock & \checkmark &  &  &  & \checkmark\\
        Malcontent & \checkmark & \checkmark & \checkmark & \checkmark &\\
        \rowcolor[HTML]{EEEEEE} OSS-Detect-Backdoor (ODB) & \checkmark & \checkmark & \checkmark & \checkmark & \\
        Packj & \checkmark & \checkmark & \checkmark & \checkmark & \\
        \rowcolor[HTML]{EEEEEE} VirusTotal & \checkmark & \checkmark &  &  &\\
        \hline
    \end{tabular}

    \label{tab:malware_detection_tools}
\end{table*}

\subsection{Linux Malware Detection Tools}
Table~\ref{tab:malware_detection_tools} lists the six open-source malware tools used in this study. Benchmarking this set of open-source tools is valuable for assessing their efficacy and reliability. In particular, we initially consulted online resources to identify Linux malware detection tools. We then refined our list by excluding tools that are no longer maintained, such as LMD~\cite{lmd}. Each of these tools has been selected due to its popularity within the community. Closed-source tools were excluded from this analysis because their underlying code is unavailable for scrutiny, and obtaining access to them requires significant time and third-party cooperation. Below, we offer a short description of each studied tool:

\begin{itemize}
    \item \textbf{\textit{Bandit4Mal}}~\cite{bandit4mal,vu2021lastpymile}: A tool designed to detect common malware patterns in Python code. It processes each file by building an Abstract Syntax Tree (AST) and runs appropriate plugins against the AST nodes. After scanning all files, it generates a report highlighting potential malware. Notably, \textit{Bandit4Mal} operates exclusively on source code.
    \item \textbf{\textit{Malcontent}}~\cite{malcontent}: A tool developed by Chainguard, it analyzes binary executables to detect malicious capabilities. It disassembles binaries to inspect them for suspicious behaviors based on over \num{16000} YARA detection rules. For instance, in the aftermath of the \textit{XZ} attack, the YARA rules underpinning \textit{Malcontent} were updated so that the tool could successfully detect this attack.  The later analysis only considers the ``High" and ``Critical" alerts. Of note, \textit{Malcontent} is able to scan both source code and Wolfi APKs.
    \item \textbf{\textit{OSS-Detect-Backdoor} (ODB for short)}~\cite{odb}: This tool is specifically designed to identify potential backdoors in open-source software projects. It scans code repositories to detect unusual code patterns. \textit{ODB} is compatible with both source code and Wolfi APKs.
    \item \textbf{\textit{Packj}}~\cite{packj}: \textit{Packj} is an open-source tool that audits packages and dependencies for security vulnerabilities and malicious code. It analyzes package metadata, source code, and dependency chains to identify potential risks. In this study, we have customized this tool to scan local source code files.
    \item \textbf{VirusTotal}~\cite{virustotal}: \textit{VirusTotal} is an online service that allows users to submit Linux samples to assess their maliciousness. It incorporates nearly \num{90} security vendors that scan samples and assign a malicious or benign label. 
    \textit{VirusTotal} is compatible with both source code and APKs.
    \item \textbf{Capslock}~\cite{capslock}: \textit{Capslock} is a tool for capability analysis of Go packages. The tool classifies the capabilities of Go packages by analyzing transitive calls to privileged standard library operations. \textit{Capslock} operates exclusively on source code.
\end{itemize}

These tools were employed to evaluate all upstream Wolfi repositories and a set of known malicious examples in Subsection~\ref{subsec:datasets}. Since not all tools produce a binary decision output (i.e., benign or malicious), some of the analyses focused on a quantitative assessment of the number of alerts. The default detection threshold was set to one, implying that if a sample triggers at least one alert, it is considered malicious.  

Only \textit{Malcontent} and \textit{Capslock} were assessed using ``over-time" datasets to evaluate their ability to detect potentially malicious changes over time. Furthermore, because it cannot be safely assumed that all Wolfi packages are benign, the analysis included a manual review of a sample of alerts for Wolfi packages. At least \num{10} alerts triggered by each scanner on Wolfi packages were manually reviewed to confirm that they were indeed benign.

\begin{table*}[!htb]
\centering
\caption{Distribution of generated alerts on the samples in the combined dataset}
\setlength{\tabcolsep}{3pt}
\begin{tabular}{@{}lS[table-format=4.2]S[table-format=6.0]S[table-format=6.2]S[table-format=6.2]S[table-format=6.2]S[table-format=6.2]S[table-format=6.2]@{}}
\hline
 & \multicolumn{1}{l}{Min} & \multicolumn{1}{l}{Max} & \multicolumn{1}{l}{Mean} & \multicolumn{1}{l}{Std Dev} & \multicolumn{1}{l}{Q25} & \multicolumn{1}{l}{Median} & \multicolumn{1}{l}{Q75} \\ \hline
\rowcolor[HTML]{EEEEEE} Bandit4Mal & 1 & 191 & 39.32 & 42.71 & 10 & 20 & 57 \\
Malcontent & 1 & 12957 & 406.17 & 1692.65 & 39.25 & 110.5 & 242.25 \\
\rowcolor[HTML]{EEEEEE} ODB & 1 & 478611 & 3765.75 & 23430.9 & 210.75 & 628.5 & 1870 \\
Packj & 1 & 5706 & 225.58 & 633.9 & 40.75 & 94 & 212 \\
\rowcolor[HTML]{EEEEEE} VirusTotal & 0 & 29 & 13.56 & 9.14 & 6 & 12 & 21 \\
 \hline
\end{tabular}
\label{tab:tool_summary}
\end{table*}

For the ``historical" source code malware samples, as a sanity check, we reviewed the malware strategies used within each ecosystem. If the strategies were ``too similar," the research team  intentionally sampled other examples within that ecosystem that employ `different' malware strategies.

    
         
\section{RQ1: Interview Key Findings}
\label{sec:results-interview}
The interview results suggest broadly similar beliefs and experiences across the maintainers of these Linux distributions, though the experience of the Wolfi maintainers appears unique in their embrace of proactive malware scanning. There are five novel findings.

First, all maintainers expressed a similar belief that the \textit{XZ} Utils attack was a turning point for each distribution. The maintainers expressed a fear of social engineering as an attack vector and worried that this type of attack was difficult, if not impossible, for these communities to defend against reliably. Consequently, interest in detecting and preventing malware in Linux distribution package repositories increased.

Second, these maintainers generally mentioned two types of pre-existing counter-malware activities. First, most maintainers mentioned the \textit{Reproducible Builds} project and their own efforts, or their fellow maintainers' efforts, to increase the number of reproducible packages within each distribution. Second, most maintainers also generally mentioned ``cryptographic" approaches that involved ensuring that packages had been signed, though participants also noted that signing packages is only a partial defense to countering an attacker bent on inserting malware.

Third, most of the participants acknowledged that there had been, after the \textit{XZ} Utils attack, more discussion within their respective communities of ``identity" checking mechanisms as a way to foil attacks like the \textit{XZ} Utils. Interestingly, the interview participants were uniformly hostile towards this line of thinking, expressing skepticism that the community norms of these distributions were compatible with such an approach and skepticism towards the efficacy of any such approach. Also, several interviewees did discuss a trade-off between fast package updates and avoiding malware, pointing out that the strength of ``slow" Linux distributions (compared to ``fast" package registries like PyPI) is in the time for review and scrutiny that staged releases allow. In short, some participants pointed towards the relatively slow release cadence of major Linux distributions as a form of passive defense against malware. Finally, several interviewees pointed out the need for efforts to ensure that, in addition to reproducible builds, that all packages be built from ``transparent sources," i.e., easily understood and thus easily scrutinized source code.

Fourth, among Linux distributions, \textit{Wolfi}, has embraced proactive malware scanning. \textit{Wolfi} has built and released an open-source tool called \textit{Malcontent} that is deployed in production within Wolfi. \textit{Malcontent} is used to generate alerts for each new package update. The difference in the alerts between each package version is analyzed; new high or critical alerts are flagged for reviewing, blocking package updates until the alert is deemed benign.

Finally, prior to the \textit{XZ} Utils attack, interviewees indicated that proactive malware scanning was rarely considered. Further, even in the aftermath of the XZ Utils attack, there is a general skepticism towards the idea, primarily due to the perceived cost of using commercial-grade scanners and concerns over who would bear the responsibility for the associated overhead of reviewing alerts. Even maintainers who express some openness about such an approach remain uncertain about where to start and how to integrate them into their workflow. Interestingly, as previously noted, Wolfi has already adopted a malware scanning approach, using \textit{Malcontent}. However, the effectiveness of \textit{Malcontent}, at least prior to this study, remains unclear.


\begin{tcolorbox}[boxrule=1.1pt]
    \textbf{Main Findings:} Linux distribution maintainers perceive a substantial and ongoing threat from malicious packages and believe that the \textit{XZ} Utils incident will not be an isolated case. Our interviews indicates that past approaches to malware detection have largely been focused on efforts related to \textit{Reproducible Builds} and cryptographic signing. Finally, with the exception of Wolfi, there has been minimal experimentation with proactive malware scanning.
\end{tcolorbox}
\section{RQ2: Malware Tools Performance}
\label{sec:results-malware}
\noindent

To compare the performance of the malware detection tools, we determine a threshold based on the number of rules or antivirus matches. In particular, the exact threshold values used in the tables in this section are set to one.
The studied tools generally have high success rates with minimal failures across most datasets, indicating good reliability for these analysis tools. Some samples cannot be processed by tools because of the tools' limitations. For example, there are six Wolfi APKs and four upstream repositories that cannot be processed by \textit{VirusTotal} as their sizes are larger than \num{650}MB. \textit{Malcontent} can process all the samples that the tool supports, while we observe a higher number of failed analyses of \textit{ODB}. \textit{Bandit4mal} fails to process only one upstream repository.

\begin{figure}[t]
\centering
\includegraphics[scale=0.5]{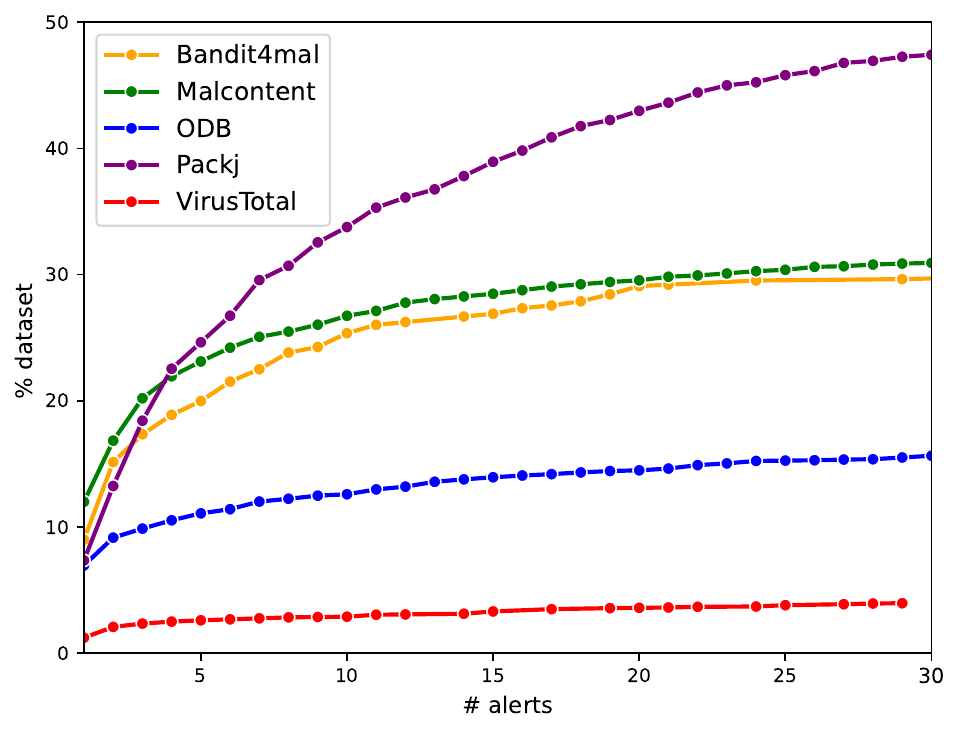}
\caption{Distribution of alerts generated by the malware scanners on the combined dataset}
\label{fig:plot_scatter}
\end{figure}

Figure~\ref{fig:plot_scatter} compares the performance of five different tools (\textit{VirusTotal}, \textit{Malcontent}, \textit{ODB}, \textit{Packj}, and \textit{Bandit4Mal}) in terms of the percentage of the dataset that generates alerts versus the total number of alerts produced by each tool. \textit{Packj} appears to generate the highest number of alerts. Table~\ref{tab:tool_summary} shows that, on average, \textit{ODB} produces the highest number of alerts, with a median of \num{628} alerts and a mean of (value missing). We observe that \textit{Malcontent} generated \num{12957} alerts for a Wolfi upstream project called \texttt{py3-azure-core}, which indicates that the tool (statement incomplete). The alerts generated by \textit{Packj} and \textit{Bandit4Mal} seem manageable by humans, as they are either specific to Python or contain more precise rules.  

Table~\ref{tab:dataset1-upstream-repos} presents a confusion matrix for several security analysis tools—Malcontent, \textit{Virustotal}, \textit{Bandit4Mal}, \textit{ODB}, and \textit{Packj} on malicious source code, indicating their performance in classifying malicious and benign files. \textit{VirusTotal} shows the best balance between identifying malicious files (True Positive) and correctly recognizing benign files (True Negative), suggesting it is the most reliable tool among those listed.  Each tool has a distinct profile in terms of its error types. For high-sensitivity tasks (e.g., minimizing False Negative), \textit{ODB} is effective, while \textit{VirusTotal} is a suitable choice for balanced accuracy. Tools like \textit{Bandit4Mal} and \textit{Packj} may benefit from further tuning to reduce their error rates in both false positives and false negatives.

\begin{table*}[!htb]
\centering
\caption{Dataset \#1 vs Benign Wolfi upstream repos}
\setlength{\tabcolsep}{6pt}
\begin{tabular}{@{}lS[table-format=4.0]S[table-format=4.0]S[table-format=4.0]S[table-format=4.0]S[table-format=4.0]S[table-format=4.0]S[table-format=4.0]S[table-format=4.0]S[table-format=4.0]S[table-format=4.0]@{}}
\hline
 & \multicolumn{2}{l}{Bandit4Mal} & \multicolumn{2}{l}{Malcontent} & \multicolumn{2}{l}{ODB} & \multicolumn{2}{l}{Packj} & \multicolumn{2}{l}{VirusTotal} \\ \hline
 & {\scriptsize Malicious} & {\scriptsize Benign} & {\scriptsize Malicious} & {\scriptsize Benign} & {\scriptsize Malicious} & {\scriptsize Benign} & {\scriptsize Malicious} & {\scriptsize Benign} & {\scriptsize Malicious} & {\scriptsize Benign} \\
 \rowcolor[HTML]{EEEEEE} {\scriptsize Malicious} & 4 & 6 & 16 & 14 & 29 & 1 & 18 & 11 & 22 & 8 \\
{\scriptsize Benign} & 286 & 605 & 934 & 922 & 1046 & 729 & 790 & 393 & 59 & 1796 \\ \hline
\end{tabular}
\label{tab:dataset1-upstream-repos}
\end{table*}

When applying to projects with injected malicious code in \textit{Dataset \#4}, Table~\ref{tab:dataset4-upstream-repos} shows that \textit{VirusTotal} appears to be the most balanced tool, maintaining a good trade-off between high true positives and low false positives. \textit{ODB}, while effective in identifying malicious files, may produce too many false positives. \textit{ODB} and \textit{Packj} demonstrate high sensitivity, as they correctly identified most malicious files. \textit{Bandit4Mal} also exhibits high sensitivity, as it did not miss any malicious files, albeit with fewer true positives. \textit{Malcontent} and \textit{ODB} exhibit notably high false positive counts, suggesting they are more conservative in classification, potentially resulting in a higher number of benign files flagged as malicious. 

\begin{table*}[!htb]
\centering
\caption{Dataset \#3 vs Wolfi upstream repositories}
\setlength{\tabcolsep}{6pt}
\begin{tabular}{@{}lS[table-format=4.0]S[table-format=4.0]S[table-format=4.0]S[table-format=4.0]S[table-format=4.0]S[table-format=4.0]S[table-format=4.0]S[table-format=4.0]S[table-format=4.0]S[table-format=4.0]@{}}
\hline
 & \multicolumn{2}{l}{Bandit4Mal} & \multicolumn{2}{l}{Malcontent} & \multicolumn{2}{l}{ODB} & \multicolumn{2}{c}{Packj} & \multicolumn{2}{l}{VirusTotal} \\ \hline
 & {\scriptsize Malicious} & {\scriptsize Benign} & {\scriptsize Malicious} & {\scriptsize Benign} & {\scriptsize Malicious} & {\scriptsize Benign} & {\scriptsize Malicious} & {\scriptsize Benign} & {\scriptsize Malicious} & {\scriptsize Benign} \\
\rowcolor[HTML]{EEEEEE} {\scriptsize Malicious} & 10 & 0 & 21 & 9 & 26 & 1 & 25 & 5 & 21 & 9 \\
{\scriptsize Benign} & 286 & 605 & 934 & 922 & 1046 & 729 & 790 & 393 & 59 & 1796 \\ \hline
\end{tabular}
\label{tab:dataset4-upstream-repos}
\end{table*}

Table~\ref{tab:dataset2-wolfi-apks} shows that when distinguishing malicious samples in \textit{Dataset \#2} and benign Wolfi APKs, \textit{VirusTotal} offers the best overall performance with high accuracy and a balanced precision-recall tradeoff. \textit{ODB} captures nearly all malicious APKs, though it misclassifies many benign APKs. \textit{Malcontent} provides moderate performance, with balanced but generally lower metrics across the board.

\begin{table}[!htb]
\centering
\caption{Dataset \#2 vs Benign Wolfi APKs}
\setlength{\tabcolsep}{3pt}
\begin{tabular}{@{}lS[table-format=4.0]S[table-format=4.0]S[table-format=4.0]S[table-format=4.0]S[table-format=4.0]S[table-format=4.0]@{}}
\hline
 & \multicolumn{2}{l}{Malcontent} & \multicolumn{2}{l}{ODB} & \multicolumn{2}{l}{Virustotal} \\ \hline
 & {\scriptsize Malicious} & {\scriptsize Benign} & {\scriptsize Malicious} & {\scriptsize Benign} & {\scriptsize Malicious} & {\scriptsize Benign} \\ \rowcolor[HTML]{EEEEEE}
{\scriptsize Malicious} & 20 & 10 & 24 & 1 & 22 & 8 \\
{\scriptsize Benign} & 376 & 1490 & 787 & 961 & 13 & 1849 \\ \hline
\end{tabular}
\label{tab:dataset2-wolfi-apks}
\end{table}

Table~\ref{tab:dataset3-wolfi-apks} shows that each tool exhibits distinct strengths and limitations in distinguishing malicious code injected into benign Wolfi APKs and legitimate ones. \textit{VirusTotal} stands out for its high accuracy and balanced metrics, making it a strong option in contexts where high overall correctness is prioritized. \textit{ODB} is highly sensitive to malicious instances, capturing the majority of them, but it suffers from a high false positive rate. \textit{Malcontent} has moderate accuracy but struggles with precision, suggesting limited reliability in its classification of malicious instances.

\begin{table}[!htb]
\centering
\caption{Dataset \#4 vs Malicious Wolfi APKs}
\setlength{\tabcolsep}{5pt}
\begin{tabular}{@{}lS[table-format=4.0]S[table-format=4.0]S[table-format=4.0]S[table-format=4.0]S[table-format=4.0]S[table-format=4.0]@{}}
\hline
 & \multicolumn{2}{l}{Malcontent} & \multicolumn{2}{l}{ODB} & \multicolumn{2}{l}{Virustotal} \\ \hline
 & {\scriptsize Malicious} & {\scriptsize Benign} & {\scriptsize Malicious} & {\scriptsize Benign} & {\scriptsize Malicious} & {\scriptsize Benign} \\
\rowcolor[HTML]{EEEEEE} {\scriptsize Malicious} & 15 & 15 & 20 & 8 & 9 & 21 \\
{\scriptsize Benign} & 376 & 1490 & 787 & 961 & 13 & 1849 \\ \hline
\end{tabular}
\label{tab:dataset3-wolfi-apks}
\end{table}

Table~\ref{tab:dataset5-wolfi-apks} shows that \textit{VirusTotal} is the most balanced and reliable of the three tools, achieving high accuracy and reasonable recall, making it suitable for cases where overall correctness is prioritized. \textit{Malcontent} has moderate accuracy but low precision and recall, suggesting it may require improvements to be practical for detecting malicious instances effectively. \textit{ODB} performs poorly across all metrics, indicating limited effectiveness in both detecting malicious instances and minimizing false positives.

\begin{table}[!htb]
\centering
\caption{Dataset \#5 vs Benign Wolfi APKs}
\setlength{\tabcolsep}{5pt}
\begin{tabular}{@{}lS[table-format=4.0]S[table-format=4.0]S[table-format=4.0]S[table-format=4.0]S[table-format=4.0]S[table-format=4.0]@{}}
\hline
 & \multicolumn{2}{l}{Malcontent} & \multicolumn{2}{l}{ODB} & \multicolumn{2}{l}{Virustotal} \\ \hline
 & {\scriptsize Malicious} & {\scriptsize Benign} & {\scriptsize Malicious} & {\scriptsize Benign} & {\scriptsize Malicious} & {\scriptsize Benign} \\
\rowcolor[HTML]{EEEEEE} {\scriptsize Malicious} & 4 & 6 & 1 & 9 & 6 & 4 \\
{\scriptsize Benign} & 376 & 1490 & 787 & 961 & 13 & 1849 \\ \hline
\end{tabular}
\label{tab:dataset5-wolfi-apks}
\end{table}

\subsection{Over time Dataset Analysis}\label{subsec:over-time}

The over time analysis in Table~\ref{tab:dataset6-overtime}, can help in understanding the consistency of classification, the evolution of detection accuracy, or the patterns in misclassification over time. In both datasets, some samples originally classified as malicious were later classified as benign (and vice versa). In Table~\ref{tab:dataset6-overtime}, the results are an aggregation across the deltas between the versions described in Section~\ref{sec:methods}. In total, we select ten projects, each with five versions. The total dataset creates \num{40} sets of alerts when taking the difference between versions. Ten sets should trigger an alert for adding malware, and the remaining \num{30} should be benign, meaning no alerts are generated. The results are recorded in a binary fashion; if any alerts are triggered, the entire analysis is counted as one for that project version difference. However, we also record the unique alerts to represent what a practitioner would see when using the tools.

Both \textit{Capslock} and \textit{Malcontent} produced a precision of \num{0.33}. \textit{Capslock} has a recall of \num{0.60}, and malcontent has a recall of \num{0.40}. However, when using the binary trigger, the analysis can be misleading. Regarding the unique alert count, \textit{Capslock} has a precision of \num{0.09} (\num{853} alerts), and \textit{Malcontent} has a precision of \num{0.28} (\num{21} alerts). These results indicate that relying on the difference in alerts between scanned software versions can still be noisy and time-consuming for practitioners to review. 

\begin{table}[!htb]
\centering
\caption{Dataset 6 - Over time Dataset Analysis}
\setlength{\tabcolsep}{5pt}
\begin{tabular}{@{}lS[table-format=4.0]S[table-format=4.0]S[table-format=4.0]S[table-format=4.0]@{}}
\hline
 & \multicolumn{2}{l}{Capslock} & \multicolumn{2}{l}{Malcontent} \\ \hline
 & {\scriptsize Malicious} & {\scriptsize Benign} & {\scriptsize Malicious} & {\scriptsize Benign} \\
\rowcolor[HTML]{EEEEEE} {\scriptsize Malicious} & 6 & 4 & 4 & 6 \\
{\scriptsize Benign} & 12 & 18 & 8 & 22 \\ \hline
\end{tabular}
\label{tab:dataset6-overtime}
\end{table}

\begin{tcolorbox}[boxrule=1.1pt]
    \textbf{Main Findings:} Among the tools, \textit{VirusTotal} proved the most reliable tool, balancing high accuracy in detecting both malicious and benign files. \textit{ODB} demonstrated strong sensitivity but exhibited numerous false positives. \textit{Malcontent} displayed moderate accuracy with lower precision, often mislabeling benign files as malicious. \textit{Bandit4Mal} and \textit{Packj} were sensitive but need tuning to reduce errors. Overall, \textit{VirusTotal} performed best in balanced detection. Using scanners in a differential setting overall reduces the amount of false positives. \textit{Malcontent} was less noisy than \textit{Capslock}, but the later captured more malware.
\end{tcolorbox}

\section{Discussion}
\label{sec:discussion}

 \subsection{Implications of Findings and Opportunities for Future Work}
The interviews make clear that maintainers generally perceive malware scanning as insufficient. This intuition was confirmed through our empirical analysis, which points to potential areas for improvement in automating malware detection. Identifying malicious software is a general challenge, but it is even more difficult in the context of open source, where curating an accurate and comprehensive dataset for evaluation purposes is particularly challenging. This difficulty may partly explain why the current malware scanners underperform in detecting malicious software.

Additionally, the maintainers prioritize other efforts as potential first lines of defense. Initiatives such as \textit{Reproducible Builds} and software provenance were considered promising. These techniques help mitigate the risk of backdoored software.

\subsection{Recommendations for improving malware detection tools}
  
\vspace{.1em}\noindent\textbf{Better malware detection rules.} The integration of the tools can benefit significantly from the quality of the rules. For example, \textit{ODB} can catch most malicious samples, but also produces a lot of false positives. As suggested by Vu et al.~\cite{vu2023bad}, some rules are better than others; therefore, further research is needed to define the severity of alerts generated by tools in association with malicious behaviors. For example, we only consider ``High" and ``Critical" alerts as indicators of maliciousness.  
\vspace{.1em}\noindent\textbf{Dynamic analysis.} The scanners evaluated in this study are static, which may not be suitable for analyzing obfuscated code. Dynamic analysis can be beneficial, especially in over-time analysis. A promising direction is to develop a dynamic analysis component (e.g., \textit{gVisor}~\cite{gvisor}) in \textit{Malcontent} to monitor new capabilities associated with malicious changes.   

\vspace{.1em}\noindent\textbf{Fine-tuning malware scanners}: Tools like \textit{Bandit4Mal} and \textit{Packj} may benefit from further tuning to reduce their rates of both false positives and false negatives. Several studies~\cite{duan2021towards, taylor2020spellbound} suggest that combining source code analysis with other types of analysis, such as metadata analysis, can help identify malicious packages more effectively.

\section{Related Work}
\label{sec:related}
This work connects two major thrusts within open-source software supply chain security: \emph{processes for detecting malicious OSS packages} and \emph{systems to protect the software supply chain}.

\subsection{Processes for Open Source Package Malware Detection \& Attack Taxonomies}

\vspace{.1em}\noindent\textbf{OSS Attack Taxonomies \& Datasets.}
There exists a myriad of work within this particular discipline. Perhaps the most visible one is that of Ohm et al.~\cite{ohm2020backstabber}, which provides a taxonomy of attack vectors
and a dataset of malicious software packages used in real-world attacks on open-source software supply chains on three
programming language package repositories (npm, PyPI, and RubyGems). Tschacher~\cite{tschacher2016typosquatting} presented a comprehensive analysis of typosquatting attacks, including the systematic generation of typosquatting package names and the publication of forks of the original packages in several open-source ecosystems. Lastly, Vu et al.~\cite{vu2020typosquatting} studied the potential impact of typosquatting attacks on PyPI packages based on the Levenshtein distance and the number of downloads of the targeted package. 
They show a large number of typosquatting candidates in PyPI that PyPI administrators should investigate. Our study investigated not only malicious packages in language-based package repositories but also their corresponding Linux binaries (APKs) built on open-source packages.

\vspace{0.1em}\noindent\textbf{Systems \& Processes for OSS Malware Detection.} Vu et al.~\cite{vu2023bad} conducted a comprehensive evaluation of three malware detection tools, two of which are being used in this paper. 
They found that these evaluated detectors had false positive rates between \qty{15}{\percent} and \qty{97}{\percent}, which was far from reaching package registry administrators’ demands (less than \qty{0.01}{\percent}). 
This evaluation, however, does not consider scanning open-source packages in Linux distributions, which have a different curation process. A similar line of work attempts to detect malicious code by studying developer behavior on version control system metadata. Notably, Gonzalez et al.~\cite{gonzalez2021anomalicious} uses commit logs and repository metadata to detect potential malicious commits. The method identifies \qty{53.3}{\percent} of malicious commits while flagging less than \qty{1}{\percent} of commits in the studied dataset. This research, however, does not investigate malicious code injected into Linux utilities and their upstream source code repositories.

\subsection{Systems to Protect the Software Supply Chain}

\vspace{0.1em}\noindent\textbf{Reproducible Builds and other software supply chain security research}
Reproducible builds~\cite{lamb2021reproducible} can help identify malicious code by allowing developers and users to verify that the source code they see is exactly what is in the final executable or software package, without any hidden alterations or malicious additions. 
Although these techniques are promising, it is time-consuming for large projects to be reproducible. 
For example, the Debian Linux distribution has worked for years to make its entire repository reproducible, and while they have made significant progress, it remains an ongoing effort due to the scale and complexity.
Our interviews highlighted that the large investment follows the developers' expectation of a security improvement~\cite{debian-reproducible}.

Fourne et al.~\cite{fourne2023s} interviewed \num{24} participants from the \textit{Reproducible-Builds} project~\cite{lamb2021reproducible}, identifying experiences that help and hinder adoption. The authors discussed the challenges of achieving Reproducible Builds, such as variability in build environments. The authors emphasize that while reproducible builds may not be easy to achieve, they are a necessary step toward a more secure and trustworthy software supply chain.

\vspace{0.1em}\noindent\textbf{Software Signing, Attestations and Provenance}
Newman et al.~\cite{newman2022sigstore} presents \textit{Sigstore}, a project that enables trusted parties to make authenticated claims about software artifacts. 
However, Sigstore does not detect malware but helps ensure that the code is from a trusted source and has not been tampered with. 
During the curation process, developers must verify the signatures of upstream sources, efforts such as Signstar~\cite{arch-signstar} are attempting to integrate these new software signing solutions into their pipelines.

To provide more transparency on software supply chain processes, Torres-Arias et al.~\cite{torres2019toto} developed \textit{in-toto}, a framework providing end-to-end security for the software supply chain of Linux distributions by enforcing detailed accountability and traceability for each step in the development and packaging processes. 
However, \textit{in-toto} does not need to recognize specific malware signatures because it focuses on verifying the integrity of each step in the software supply chain. 
However, \textit{in-toto} may be a tool of choice to communicate malware scanning results across different Linux distributions as we discussed in Section~\ref{sec:discussion}.
Other systems, such as \textit{Contour}~\cite{al2018contour} and \textit{CHAINIAC}~\cite{nikitin2017chainiac} may also be viable alternatives for this goal.

\subsection{Qualitative Research on Linux Communities}

Qualitative research into Linux communities has focused on understanding the motivations~\cite{zhao1999qualitative, hertel2003motivation}, communications, retention factors, and challenges~\cite{hughes1996lj} that contributors experience.  Wermke et al.~\cite{wermke2022committed} investigated security challenges encountered by contributors such as ``hypocrite commits" or malicious commits. 
Similarly, work by Lin~\cite{lin2023vulnerability} explored the practices in Linux distributions when managing vulnerabilities reported on the Linux Kernel. However, these works do not investigate proactive malware scanning practices on the Linux distributions. 
\section{Threats to Validity}
\label{sec:validity}

There are both possible internal and external threats to the validity of this research.

\subsection{Internal Threats to Validity}
\vspace{.1em}\noindent\textbf{Data Labeling Errors.}
The benign dataset relies on the assumption that the packages and their upstream repositories are benign. This assumption has not been thoroughly validated. We did not undertake the monumental task of manually verifying that all binaries and their upstream source code are, in fact, benign. Consequently, there is a risk that some samples labeled as benign may be malicious. If such mislabeled samples exist, they would incorrectly inflate the false positive rates. Although this possibility cannot be entirely ruled out, we considered it a safe assumption that only a handful of packages might fall into this category.

\vspace{.1em}\noindent\textbf{Sampling Bias in Malware Selection.}
The malware samples used in this study are unlikely to represent the full spectrum of threats encountered in real-world environments. The malicious datasets include, for instance, some heavily studied malware instances. As a result, existing malware tools may already be relatively well-adapted to detect these threats. This sampling bias could lead to benchmarking results that overestimate the effectiveness of the evaluated tools.

\vspace{.1em}\noindent\textbf{The current evaluation treats false positives and false negatives equally.} Our interviews with Linux distro maintainers, plus our reading of Vu et al.'s paper ``Bad Snakes"~\cite{vu2023bad} and its interviews of PyPI maintainers, suggest that package repository maintainers are, perhaps surprisingly, quite sensitive to high false positive rates. These volunteer communities are reluctant to spend their time sifting through findings that are false positives, despite the danger of zero-day malware and new variants. Furthermore, while focusing on differences between releases is a theoretically appealing approach, the quantitative results in Subsection~\ref{subsec:over-time} suggest that such a ``version diff" approach might not be as empirically useful as theory would predict.

\subsection{External Threats to Validity}
\vspace{.1em}\noindent\textbf{Selection Bias in Interview Samples.}
Our interview sample may be affected by selection bias or survivor bias. The sample predominantly consists of experienced maintainers who have been active in the field for a significant period. This selection may have excluded newer maintainers or those involved in less mainstream or ``fringe" projects. Consequently, the perspectives gathered from the interviews may not fully represent the experiences of all maintainers.

\vspace{.1em}\noindent\textbf{Evolving and Diverse Malware Threats.}
The malware dataset used in this study is primarily composed of historical samples, including those from sources like the Backstabber's Knife Collection. While these samples provide valuable insights into well-known attack patterns, they may not adequately capture the evolving and increasingly sophisticated nature of malware.

\vspace{.1em}\noindent\textbf{Impact of Scanner Evolution and Updates.}
The rapid evolution of scanner technologies introduces another potential limitation. As new malware variants are discovered and scanners release updated detection signatures and algorithms, the performance of these tools could improve significantly. Therefore, the results presented in this study may not accurately reflect the current capabilities of these scanners, particularly if substantial updates to detection engines have occurred since the time of writing.

\section{Conclusion}
\label{sec:conclusion}
Malware prevention is an increasingly significant concern for Linux distributions and their maintainers. Through interviews, we identified that most Linux distributions, excluding Wolfi, do not perform active malware scanning. The Wolfi maintainers' use of \textit{Malcontent} is helping to pioneer this active scanning approach by combining program capability analysis with malware detection techniques to identify Linux package malware threats.

By building a set of Linux package malware benchmark datasets and scanning them with existing malware scanners, we found that these tools exhibit high false positive rates and struggle to accurately identify malware. Our analysis revealed that none of the tools achieved detection rates above \qty{70}{\percent}. As a result, these tools are unlikely to be widely adopted by other Linux distributions and are likely to frustrate maintainers who attempt to use them.

In summary, this paper emphasizes the urgent need to improve malware detection within Linux package repositories. Our ultimate hope is that the next time the person or group (or even machine) behind GitHub user \textit{JiaT75} attempts to backdoor a popular open-source project, these tools or improved versions of them, or entirely new tools will detect the attack so swiftly that it becomes a non-event, failing to garner attention in the news. Here is to hoping.

\bibliographystyle{IEEEtran}
\bibliography{refs}

\begin{thebibliography}{10}
\providecommand{\url}[1]{#1}
\csname url@samestyle\endcsname
\providecommand{\newblock}{\relax}
\providecommand{\bibinfo}[2]{#2}
\providecommand{\BIBentrySTDinterwordspacing}{\spaceskip=0pt\relax}
\providecommand{\BIBentryALTinterwordstretchfactor}{4}
\providecommand{\BIBentryALTinterwordspacing}{\spaceskip=\fontdimen2\font plus
\BIBentryALTinterwordstretchfactor\fontdimen3\font minus \fontdimen4\font\relax}
\providecommand{\BIBforeignlanguage}[2]{{%
\expandafter\ifx\csname l@#1\endcsname\relax
\typeout{** WARNING: IEEEtran.bst: No hyphenation pattern has been}%
\typeout{** loaded for the language `#1'. Using the pattern for}%
\typeout{** the default language instead.}%
\else
\language=\csname l@#1\endcsname
\fi
#2}}
\providecommand{\BIBdecl}{\relax}
\BIBdecl

\bibitem{xz}
A.~S.~I. Group, ``Xz utils backdoor - everything you need to know, and what you can do.'' \url{https://www.akamai.com/blog/security-research/critical-linux-backdoor-xz-utils-discovered-what-to-know}, 2024, accessed: 2024-11-14.

\bibitem{goodin2024}
\BIBentryALTinterwordspacing
D.~Goodin, ``\BIBforeignlanguage{en-US}{What we know about the xz utils backdoor that almost infected the world},'' Apr. 2024. [Online]. Available: \url{https://arstechnica.com/security/2024/04/what-we-know-about-the-xz-utils-backdoor-that-almost-infected-the-world/}
\BIBentrySTDinterwordspacing

\bibitem{duan2021towards}
R.~Duan, O.~Alrawi, R.~P. Kasturi, R.~Elder, B.~Saltaformaggio, and W.~Lee, ``Towards measuring supply chain attacks on package managers for interpreted languages,'' in \emph{Proc. of NDSS'21}, 2021.

\bibitem{ohm2020backstabber}
M.~Ohm, H.~Plate, A.~Sykosch, and M.~Meier, ``Backstabber’s knife collection: A review of open source software supply chain attacks,'' in \emph{International Conference on Detection of Intrusions and Malware, and Vulnerability Assessment}.\hskip 1em plus 0.5em minus 0.4em\relax Springer, 2020, pp. 23--43.

\bibitem{vu2020typosquatting}
D.-L. Vu, I.~Pashchenko, F.~Massacci, H.~Plate, and A.~Sabetta, ``Typosquatting and combosquatting attacks on the python ecosystem,'' in \emph{2020 IEEE European Symposium on Security and Privacy Workshops (EuroS\&PW)}, 2020.

\bibitem{its_2016}
\BIBentryALTinterwordspacing
I.~Z. Schlueter, ``npm blog archive: kik, left-pad, and npm,'' Mar. 2016. [Online]. Available: \url{https://blog.npmjs.org/post/141577284765/kik-left-pad-and-npm}
\BIBentrySTDinterwordspacing

\bibitem{arvanitis_2022}
\BIBentryALTinterwordspacing
I.~Arvanitis, G.~Ntousakis, S.~Ioannidis, and N.~Vasilakis, ``A systematic analysis of the event-stream incident,'' in \emph{Proceedings of the 15th European Workshop on Systems Security}, ser. EuroSec '22.\hskip 1em plus 0.5em minus 0.4em\relax New York, NY, USA: Association for Computing Machinery, 2022, p. 22–28. [Online]. Available: \url{https://doi.org/10.1145/3517208.3523753}
\BIBentrySTDinterwordspacing

\bibitem{Nitansha_2024}
\BIBentryALTinterwordspacing
N.~Bansal and S.~Scott, ``\BIBforeignlanguage{en-US}{The 5x5—the xz backdoor: Trust and open source software},'' May 2024. [Online]. Available: \url{https://dfrlab.org/2024/05/01/the-5x5-the-xz-backdoor-trust-and-open-source-software/}
\BIBentrySTDinterwordspacing

\bibitem{thompson1984reflections}
K.~Thompson, ``Reflections on trusting trust,'' \emph{Communications of the ACM}, vol.~27, no.~8, pp. 761--763, 1984.

\bibitem{wheeler2005countering}
D.~A. Wheeler, ``Countering trusting trust through diverse double-compiling,'' in \emph{21st Annual Computer Security Applications Conference (ACSAC'05)}.\hskip 1em plus 0.5em minus 0.4em\relax IEEE, 2005, pp. 13--pp.

\bibitem{lamb2021reproducible}
C.~Lamb and S.~Zacchiroli, ``Reproducible builds: Increasing the integrity of software supply chains,'' \emph{IEEE Software}, vol.~39, no.~2, pp. 62--70, 2021.

\bibitem{torres2019toto}
S.~Torres-Arias, H.~Afzali, T.~K. Kuppusamy, R.~Curtmola, and J.~Cappos, ``in-toto: Providing farm-to-table guarantees for bits and bytes,'' in \emph{28th USENIX Security Symposium (USENIX Security 19)}, 2019, pp. 1393--1410.

\bibitem{geer2020good}
D.~Geer, B.~Tozer, and J.~S. Meyers, ``For good measure: Counting broken links: A quant’s view of software supply chain security,'' \emph{USENIX; Login}, vol.~45, no.~4, 2020.

\bibitem{Office_2022}
\BIBentryALTinterwordspacing
U.~S. G.~A. Office, ``\BIBforeignlanguage{en}{Cybersecurity: Federal response to solarwinds and microsoft exchange incidents | u.s. gao},'' Feb. 2022. [Online]. Available: \url{https://www.gao.gov/products/gao-22-104746}
\BIBentrySTDinterwordspacing

\bibitem{cisa}
\BIBentryALTinterwordspacing
CISA, ``\BIBforeignlanguage{en}{Advanced persistent threat compromise of government agencies, critical infrastructure, and private sector organizations},'' Apr. 2021. [Online]. Available: \url{https://www.cisa.gov/news-events/cybersecurity-advisories/aa20-352a}
\BIBentrySTDinterwordspacing

\bibitem{zahan2024s3c2}
N.~Zahan, Y.~Acar, M.~Cukier, W.~Enck, C.~K{\"a}stner, A.~Kapravelos, D.~Wermke, and L.~Williams, ``S3c2 summit 2023-11: Industry secure supply chain summit,'' \emph{arXiv preprint arXiv:2408.16529}, 2024.

\bibitem{xia2023empirical}
B.~Xia, T.~Bi, Z.~Xing, Q.~Lu, and L.~Zhu, ``An empirical study on software bill of materials: Where we stand and the road ahead,'' in \emph{2023 IEEE/ACM 45th International Conference on Software Engineering (ICSE)}.\hskip 1em plus 0.5em minus 0.4em\relax IEEE, 2023, pp. 2630--2642.

\bibitem{zahan2023sbom}
N.~Zahan, E.~Lin, M.~Tamanna, W.~Enck, and L.~Williams, ``Software bills of materials are required. are we there yet?'' \emph{IEEE Security \& Privacy}, vol.~21, no.~2, pp. 82--88, 2023.

\bibitem{newman2022sigstore}
Z.~Newman, J.~S. Meyers, and S.~Torres-Arias, ``Sigstore: Software signing for everybody,'' in \emph{Proceedings of the 2022 ACM SIGSAC Conference on Computer and Communications Security}, 2022, pp. 2353--2367.

\bibitem{merrill2023speranza}
K.~Merrill, Z.~Newman, S.~Torres-Arias, and K.~R. Sollins, ``Speranza: Usable, privacy-friendly software signing,'' in \emph{Proceedings of the 2023 ACM SIGSAC Conference on Computer and Communications Security}, 2023, pp. 3388--3402.

\bibitem{slsa}
\BIBentryALTinterwordspacing
``\BIBforeignlanguage{en}{Supply-chain levels for software artifacts}.'' [Online]. Available: \url{https://slsa.dev/}
\BIBentrySTDinterwordspacing

\bibitem{moriconi2023reflections}
F.~Moriconi, A.~I. Neergaard, L.~Georget, S.~Aubertin, and A.~Francillon, ``Reflections on trusting docker: Invisible malware in continuous integration systems,'' in \emph{2023 IEEE Security and Privacy Workshops (SPW)}.\hskip 1em plus 0.5em minus 0.4em\relax IEEE, 2023, pp. 219--227.

\bibitem{vu2023bad}
D.-L. Vu, Z.~Newman, and J.~S. Meyers, ``Bad snakes: Understanding and improving python package index malware scanning,'' in \emph{2023 IEEE/ACM 45th International Conference on Software Engineering (ICSE)}.\hskip 1em plus 0.5em minus 0.4em\relax IEEE, 2023, pp. 499--511.

\bibitem{ohm2022feasibility}
M.~Ohm, F.~Boes, C.~Bungartz, and M.~Meier, ``On the feasibility of supervised machine learning for the detection of malicious software packages,'' in \emph{Proceedings of the 17th International Conference on Availability, Reliability and Security}, 2022, pp. 1--10.

\bibitem{zahan2024shifting}
N.~Zahan, P.~Burckhardt, M.~Lysenko, F.~Aboukhadijeh, and L.~Williams, ``Shifting the lens: Detecting malware in npm ecosystem with large language models,'' \emph{arXiv preprint arXiv:2403.12196}, 2024.

\bibitem{ladisa2023sok}
P.~Ladisa, H.~Plate, M.~Martinez, and O.~Barais, ``Sok: Taxonomy of attacks on open-source software supply chains,'' in \emph{2023 IEEE Symposium on Security and Privacy (SP)}.\hskip 1em plus 0.5em minus 0.4em\relax IEEE, 2023, pp. 1509--1526.

\bibitem{yin2015qualitative}
R.~K. Yin, \emph{Qualitative research from start to finish}.\hskip 1em plus 0.5em minus 0.4em\relax Guilford publications, 2015.

\bibitem{wolfi}
\BIBentryALTinterwordspacing
Chainguard, ``\BIBforeignlanguage{en}{Wolfi os},'' 2024. [Online]. Available: \url{https://github.com/wolfi-dev/os}
\BIBentrySTDinterwordspacing

\bibitem{GitHubAPI}
\BIBentryALTinterwordspacing
{GitHub Inc.}, ``\BIBforeignlanguage{en}{{GitHub API Documentation}},'' 2025. [Online]. Available: \url{https://docs.github.com/en/rest}
\BIBentrySTDinterwordspacing

\bibitem{wang2025malpacdetector}
J.~Wang, Z.~Li, J.~Qu, D.~Zou, S.~Xu, Z.~Xu, Z.~Wang, and H.~Jin, ``Malpacdetector: An llm-based malicious npm package detector,'' \emph{IEEE Transactions on Information Forensics and Security}, 2025.

\bibitem{melange}
\BIBentryALTinterwordspacing
{Chainguard}, ``\BIBforeignlanguage{en}{melange},'' 2023, accessed: 2024-11-14. [Online]. Available: \url{https://github.com/chainguard-dev/melange}
\BIBentrySTDinterwordspacing

\bibitem{linux-malware}
\BIBentryALTinterwordspacing
{Timb Machine}, ``\BIBforeignlanguage{en}{Linux malware},'' 2023, accessed: 2024-11-14. [Online]. Available: \url{https://github.com/timb- machine/linux-malware}
\BIBentrySTDinterwordspacing

\bibitem{coldfire}
\BIBentryALTinterwordspacing
R.~C. Labs, ``\BIBforeignlanguage{en}{Coldfire: Golang malware development librarywolfi os},'' 2020. [Online]. Available: \url{https://github.com/redcode-labs/Coldfire}
\BIBentrySTDinterwordspacing

\bibitem{lmd}
\BIBentryALTinterwordspacing
R.~MacDonald, ``\BIBforeignlanguage{en}{Linux malware detect},'' 2025. [Online]. Available: \url{https://github.com/rfxn/linux-malware-detect}
\BIBentrySTDinterwordspacing

\bibitem{bandit4mal}
\BIBentryALTinterwordspacing
lyvd, ``\BIBforeignlanguage{en}{Badit4mal},'' 2024. [Online]. Available: \url{https://github.com/lyvd/bandit4mal}
\BIBentrySTDinterwordspacing

\bibitem{vu2021lastpymile}
D.-L. Vu, F.~Massacci, I.~Pashchenko, H.~Plate, and A.~Sabetta, ``Lastpymile: identifying the discrepancy between sources and packages,'' in \emph{Proceedings of the 29th ACM Joint Meeting on European Software Engineering Conference and Symposium on the Foundations of Software Engineering}, 2021, pp. 780--792.

\bibitem{malcontent}
\BIBentryALTinterwordspacing
Chainguard, ``\BIBforeignlanguage{en}{Malcontent},'' 2024. [Online]. Available: \url{https://github.com/chainguard-dev/malcontent}
\BIBentrySTDinterwordspacing

\bibitem{odb}
\BIBentryALTinterwordspacing
Microsoft, ``\BIBforeignlanguage{en}{Oss detect backdoor},'' 2024. [Online]. Available: \url{https://github.com/microsoft/OSSGadget/tree/main/src/oss-detect-backdoor}
\BIBentrySTDinterwordspacing

\bibitem{packj}
\BIBentryALTinterwordspacing
Ossilate, ``\BIBforeignlanguage{en}{Packj},'' 2024. [Online]. Available: \url{https://github.com/ossillate-inc/packj}
\BIBentrySTDinterwordspacing

\bibitem{virustotal}
\BIBentryALTinterwordspacing
VirusTotal, ``\BIBforeignlanguage{en}{Virustotal},'' 2024. [Online]. Available: \url{https://www.virustotal.com/}
\BIBentrySTDinterwordspacing

\bibitem{capslock}
\BIBentryALTinterwordspacing
Google, ``\BIBforeignlanguage{en}{Capslock},'' 2024. [Online]. Available: \url{https://github.com/google/capslock}
\BIBentrySTDinterwordspacing

\bibitem{gvisor}
T.~gVisor Authors, ``The container security platform,'' \url{https://github.com/google/gvisor}, accessed: 2024-11-14.

\bibitem{taylor2020spellbound}
M.~Taylor, R.~K. Vaidya, D.~Davidson, L.~De~Carli, and V.~Rastogi, ``Spellbound: Defending against package typosquatting,'' \emph{arXiv preprint arXiv:2003.03471}, 2020.

\bibitem{tschacher2016typosquatting}
N.~P. Tschacher, ``Typosquatting in programming language package managers,'' Ph.D. dissertation, Universit{\"a}t Hamburg, Fachbereich Informatik, 2016.

\bibitem{gonzalez2021anomalicious}
D.~Gonzalez, T.~Zimmermann, P.~Godefroid, and M.~Sch{\"a}fer, ``Anomalicious: Automated detection of anomalous and potentially malicious commits on github,'' in \emph{2021 IEEE/ACM 43rd International Conference on Software Engineering: Software Engineering in Practice (ICSE-SEIP)}.\hskip 1em plus 0.5em minus 0.4em\relax IEEE, 2021, pp. 258--267.

\bibitem{debian-reproducible}
\BIBentryALTinterwordspacing
Debian, ``\BIBforeignlanguage{en}{Reproducible builds},'' 2025. [Online]. Available: \url{https://wiki.debian.org/ReproducibleBuilds}
\BIBentrySTDinterwordspacing

\bibitem{fourne2023s}
M.~Fourn{\'e}, D.~Wermke, W.~Enck, S.~Fahl, and Y.~Acar, ``It’s like flossing your teeth: On the importance and challenges of reproducible builds for software supply chain security,'' in \emph{2023 IEEE Symposium on Security and Privacy (SP)}.\hskip 1em plus 0.5em minus 0.4em\relax IEEE, 2023, pp. 1527--1544.

\bibitem{arch-signstar}
A.~Linux, ``A secure enclave signing solution,'' \url{https://gitlab.archlinux.org/archlinux/signstar}, accessed: 2024-11-14.

\bibitem{al2018contour}
M.~Al-Bassam and S.~Meiklejohn, ``Contour: A practical system for binary transparency,'' in \emph{Data Privacy Management, Cryptocurrencies and Blockchain Technology: ESORICS 2018 International Workshops, DPM 2018 and CBT 2018, Barcelona, Spain, September 6-7, 2018, Proceedings 13}.\hskip 1em plus 0.5em minus 0.4em\relax Springer, 2018, pp. 94--110.

\bibitem{nikitin2017chainiac}
K.~Nikitin, E.~Kokoris-Kogias, P.~Jovanovic, N.~Gailly, L.~Gasser, I.~Khoffi, J.~Cappos, and B.~Ford, ``$\{$CHAINIAC$\}$: Proactive $\{$Software-Update$\}$ transparency via collectively signed skipchains and verified builds,'' in \emph{26th USENIX Security Symposium (USENIX Security 17)}, 2017, pp. 1271--1287.

\bibitem{zhao1999qualitative}
\BIBentryALTinterwordspacing
H.~Zhao, ``A qualitative study of the linux open source community,'' 1999. [Online]. Available: \url{https://cdr.lib.unc.edu/concern/masters_papers/v405sf00g}
\BIBentrySTDinterwordspacing

\bibitem{hertel2003motivation}
G.~Hertel, S.~Niedner, and S.~Herrmann, ``Motivation of software developers in open source projects: an internet-based survey of contributors to the linux kernel,'' \emph{Research policy}, vol.~32, no.~7, pp. 1159--1177, 2003.

\bibitem{hughes1996lj}
P.~Hughes and G.~Shurleff, ``Lj interviews linus torvalds,'' \emph{Linux Journal}, vol. 1996, no. 29es, pp. 1--es, 1996.

\bibitem{wermke2022committed}
D.~Wermke, N.~W{\"o}hler, J.~H. Klemmer, M.~Fourn{\'e}, Y.~Acar, and S.~Fahl, ``Committed to trust: A qualitative study on security \& trust in open source software projects,'' in \emph{2022 IEEE symposium on Security and Privacy (SP)}.\hskip 1em plus 0.5em minus 0.4em\relax IEEE, 2022, pp. 1880--1896.

\bibitem{lin2023vulnerability}
J.~Lin, H.~Zhang, B.~Adams, and A.~E. Hassan, ``Vulnerability management in linux distributions: An empirical study on debian and fedora,'' \emph{Empirical Software Engineering}, vol.~28, no.~2, p.~47, 2023.

\end{thebibliography}
\end{document}